\documentclass[graybox]{svmult}

\usepackage{mathptmx}       
\usepackage{helvet}         
\usepackage{courier}        
\usepackage{type1cm}        
\usepackage{makeidx}         
\usepackage{graphicx}        
\usepackage{multicol}        
\usepackage[bottom]{footmisc}

\usepackage[numbers]{natbib}

\makeindex             

\newcommand{\event}[2][]{{#2}_\mathrm{event}{#1}}
\newcommand{\normal}[2][]{\left<{#2}_\mathrm{normal}{#1}\right>}
\newcommand{\avg}[1]{\left<{#1}\right>}

\def\rc{r_\mathrm{c}}
\def\fmid{f_\mathrm{mid}}

\def\tstart{t_\mathrm{start}}
\def\tstop{t_\mathrm{stop}}

\newcommand{\lett}[1]{(\textbf{#1})}

\begin{document}

\title*{Information spreading during emergencies and anomalous events}
\titlerunning{Information spreading during emergencies and anomalous events}
\author{James P.~Bagrow}
\institute{James P.~Bagrow \at Mathematics \& Statistics, Vermont Complex Systems Center, University of Vermont, Burlington, VT 05405, USA \email{james.bagrow@uvm.edu}
}
%
%
\maketitle

\abstract{
The most critical time for information to spread is in the aftermath of a serious emergency, crisis, or disaster.
Individuals affected by such situations can now turn to an array of communication channels, from mobile phone calls and text messages to social media posts, when alerting social ties. 
These channels drastically improve the speed of information in a time-sensitive event, and
provide extant records of human dynamics during and afterward the event.
Retrospective analysis of such anomalous events provides researchers with a class of ``found experiments'' that may be used to better understand social spreading.
In this chapter, we study information spreading due to a number of emergency events, including the Boston Marathon Bombing and a plane crash at a western European airport. We also contrast the different information which may be gleaned by social media data compared with mobile phone data and we estimate the rate of anomalous events in a mobile phone dataset using a proposed anomaly detection method.
}

\section{Introduction}
\label{sec:intro}

Social networks are characterized both by their topological properties and by the dynamics they facilitate. The social spread of information is one of the most important of these dynamics~\cite{rogers2010diffusion}. 
Information spreading in the real world has been well studied. For example, Granovetter studied how individuals use their social ties to learn about new job opportunities~\cite{granovetter_weakties_1973}. 
Modern datasets such as social media and mobile phones have provided large-scale followups and confirmation to this seminal work~\cite{Onnela01052007}.

However, most research on social spreading has been limited to understanding the ordinary, day-to-day dynamics. 
Anomalous and extreme situations, such as information spreading in the wake of an emergency or disaster, has not received as much attention~\cite{bagrowDisaster2011pone}. Yet with appropriate data these situations provide a context by which researchers can better understand social networks and spreading phenomena. 
When an event occurs, a large amount of activity is generated, and this activity is all focused on that one event, leading to a strong and cohesive signal. Moreover, 
latent portions of the social network are likely to be activated, providing researchers a new view of the underlying social system, and there is no clearer indicator of the importance of a social tie than someone in the middle of an emergency or its aftermath choosing to reach out and communicate with that tie. 

In this chapter, we discuss how natural and technological emergency and disaster events can be used to better understand social systems, human dynamics, and the spread of information and misinformation. 
Evidence supports a long-term increase in the frequency and severity of such events~\cite{em2010ofda,eshghi2008disasters}, with driving factors including climate change and population growth. 
Every effort should be made to prevent human and technological disasters, but when they inevitably occur it is important to glean as much useful information from them as possible, not only for scientific understanding but also to improve our response to future events and save lives.

The rest of this chapter is organized as follows.
In Sec.~\ref{sec:background} we describe the history of the sociology of disasters, and summarize research on using social media and telecommunications data to better understand information spreading during emergencies and disasters.
In Sec~\ref{sec:twitterbostonmarathonbombing} we provide a case study of activity on Twitter related to the Boston Marathon Bombings.
In Sec~\ref{sec:mobilephonesemergencies} we summarize research on measuring activities in the wake of emergencies using mobile phone data taken from a country in Western Europe.
In Sec.~\ref{sec:miningemergencies} we introduce an algorithm to detect unusual call activity in this country-wide mobile phone data, use it to detect 340 anomalies during a six month period, and define statistics to characterize the properties of these emergencies and how emergency and non-emergency events (such as music festivals) differ.
We conclude with a discussion in Sec.~\ref{sec:discussion}.

\section{Background}
\label{sec:background}

The study of the social response to emergencies, crises, and disasters has a long 
and fruitful history within the field of sociology, much of it due to E.L.~Quarentelli, Russell Dynes, and J.~Eugene Haas, who founded the Disaster Research Center at Ohio State University and
pioneered the field of disaster sociology~\cite{quarantelli1954nature,quarantelli1977response,quarantelli1978disasters,quarantelli1988disaster,quarantelli2005disaster,kennedy2009handbook}.
This work, strongly influenced by the aftermath of World War II and the then-current climate of the Cold War, focused on organized behavior and emergent activity both during disasters and in their aftermaths~\cite{thierney2007margins}. Other pioneering work included case studies of disasters, such as panics and stampedes at large rock concerts~\cite{johnson1987panic}.
Emergencies are complicated events, however, and the way individuals react to them is challenging to study.
Indeed, even basic questions, like how much panic occurs or does not occur among individuals experiencing an emergency, is a contested area of research~\cite{fischer1998response}.

A primary focus of disaster sociology has been understanding and improving the organizational aspects of the response to a disaster. Communication problems between competing and overlapping government agencies have hampered first responders in many large-scale emergencies, including highly unpredictable situations like the 9/11 terrorist attacks and more predictable situations such as the landfall of Hurricane Katrina~\cite{lind2008brokerage}.
Researchers have studied how organizations such as first responders use communication technology, how their use of that technology has adapted to changes and modernizations~\cite{merchant2011integrating}, and how and why such organizations either under-perform or can improve in their ability to efficiently and effectively respond to emergencies and disasters.

Since the pioneering work on disaster sociology,
the rise of mobile phones, smartphones, and online social media have reshaped human communications. 
Individuals can now remain in constant contact with social ties if they choose, and can quickly broadcast to a group of online followers almost anywhere. And these broadcasts can quickly go ``viral,'' spreading very rapidly online.
Social media such as Facebook and Twitter have played key roles in recent emergency situations
~\cite{merchant2011integrating,schultz2011medium,wang2014social}.

Recent work has studied how Twitter posts spread in the event of emergencies~\cite{hughes2009twitter,wang2014social}. 
Twitter is a popular microblogging platform where users can post short messages called tweets to their online followers, as well as repost or forward other tweets by ``retweeting'' them. 
Some tweets will become heavily retweeted, leading to cascades. Some tweets are geotagged, containing the geographic coordinates of the tweet poster when the tweet was made.
Twitter posts are public and available through APIs to researchers, providing a wealth of text and activity data.
For example, Sakaki \emph{et al.} studied Twitter activity in the wake of a major earthquake, showing that tweets can be used to detect an earthquake in real-time~\cite{sakaki2010earthquake}.
Other work has studied how information (and misinformation) spreads during and after events including the Deepwater Horizon oil spill~\cite{spiro2012rumoring}, wildfires and floods~\cite{vieweg2010microblogging}, Hurricane Sandy~\cite{chatfield2014sandy}, the 2010 Haiti Earthquake~\cite{muralidharan2011hope,oh2010exploration}, and the Boston Marathon Bombing~\cite{starbird2014rumors,tapia2014run}.
Twitter is also used by government organizations such as first responders and by NGOs such as aid providers and relief organizations. Researchers have studied how these organizations use Twitter to
spread information and deal with rumors and misinformation~\cite{sutton2014terse,sutton2014warning}.

Another avenue for data on emergencies and disasters is mobile phone records, specifically voice calls and text messages\footnote{Although smartphone texting apps such as WhatsApp, Facebook Messenger, WeChat, Signal,  SnapChat, Line, Apple Messages, etc.\ are now blurring the line between mobile phone SMS texts and online social media.}. Unlike social media, these are generally not intended for broadcasting content to a group of followers, but are instead a specific, often one-on-one, communication medium. This activity is also not mixed with news media usage in the way that most journalistic organizations now rely heavily on social media. 
This one-on-one nature allows mobile phone data to more accurately capture individual social behavior and communication intent.

Researchers have studied mobile phone and smartphone activity in the wake of emergency events~\cite{raento2009smartphones,bagrowDisaster2011pone,gao2014quantifying}. 
Kapoor, Eagle, and Horvitz used mobile phone records in Africa to show that phone communications can act as early warning signals for an earthquake, and proposed an algorithm that can accurately pinpoint the epicenter of the earthquake~\cite{kapoor2010people}. 
Bagrow, Wang, and Barab\'asi~\cite{bagrowDisaster2011pone} and Gao \emph{et al.}~\cite{gao2014quantifying} used mobile phone records to study a number of emergency events occurring in a country in western Europe, including a bombing and a plane crash. These records capture the temporal and spatial localization of the event from the spike in call volume immediately following the event, as well as the social propagation of information (see also Secs.~\ref{sec:mobilephonesemergencies} and \ref{sec:miningemergencies} and Figs.~\ref{fig:plosOneTemporalSpatial}--\ref{fig:recurrenceplot}). 
While mobile phone data are less readily available for researchers than public social media activity, we propose that it is an invaluable source of information parallel to social media.

\section{Twitter during and after the Boston Marathon Bombing}
\label{sec:twitterbostonmarathonbombing}

As an example demonstrating how emergency events provide a window into human dynamics, we performed a small case study of Twitter social media activity following the Boston Marathon Bombing. 

\subsection{Background}
The Boston Marathon Bombing occurred on April 15, 2013 at 14:49 local time. Two improvised explosive devices exploded in a crowd near the Boston Marathon's finish line, killing three and injuring 264~\cite{kotz2013injury}. 
A manhunt soon unfolded, and on April 18 the FBI released photos of two suspects, Chechen-American brothers Dzhokhar Tsarnaev and Tamerlan Tsarnaev. That evening the brothers shot and killed a police officer, kidnapped a man and stole in his car, and had a shootout with police during which Tamerlan Tsarnaev was  killed. The next day on April 19, Dzhokhar Tsarnaev was shot and arrested at 20:42. A police officer wounded in the April 18 shootout died the following year. Dzhokhar Tsarnaev was convicted of multiple crimes and sentenced to death in 2015~\cite{morrison2015timeline}.

There was much related activity on Twitter in the immediate aftermath of the bombing and throughout the period of heightened uncertainty between the bombing and the capture of Dzhokhar Tsarnaev. In fact, Dzhokhar Tsarnaev himself tweeted multiple times between April 15 and April 19~\cite{morrison2015timeline}.
Further, much misinformation and rumor propagated online, including online groups making false allegations against a missing college student~\cite{tapia2014run}.

\subsection{Information and rumors on Twitter}

\begin{figure}[t]
\includegraphics[width=0.85\textwidth]{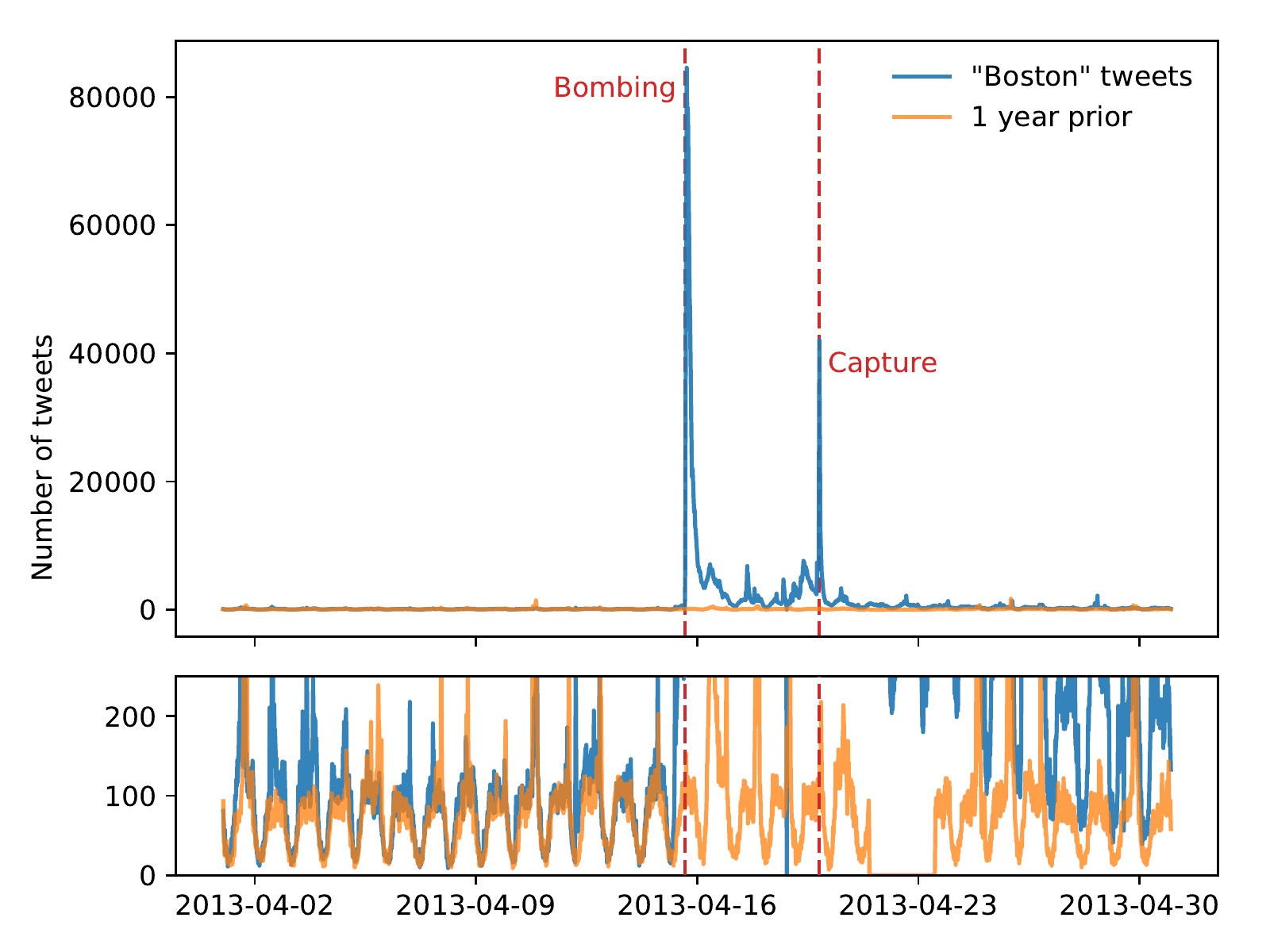}
\caption{Twitter activity surrounding the Boston Marathon Bombing.
Shown are counts of tweets containing ``boston'' (case-insensitive) during April, 2013, compared with one year earlier.
The bombing on April 15 is clearly visible, as is the capture of Dzhokhar Tsarnaev on April 19.
The year-over-year regularity of the time series before the bombing is evident in the lower plot, which shows the same time series but with a tighter range. 
Year-over-year deviations before the bombing are primarily due to sporting events.
}
\label{fig:numBostonTweets}
\end{figure}

\begin{figure}
\includegraphics[width=0.85\textwidth]{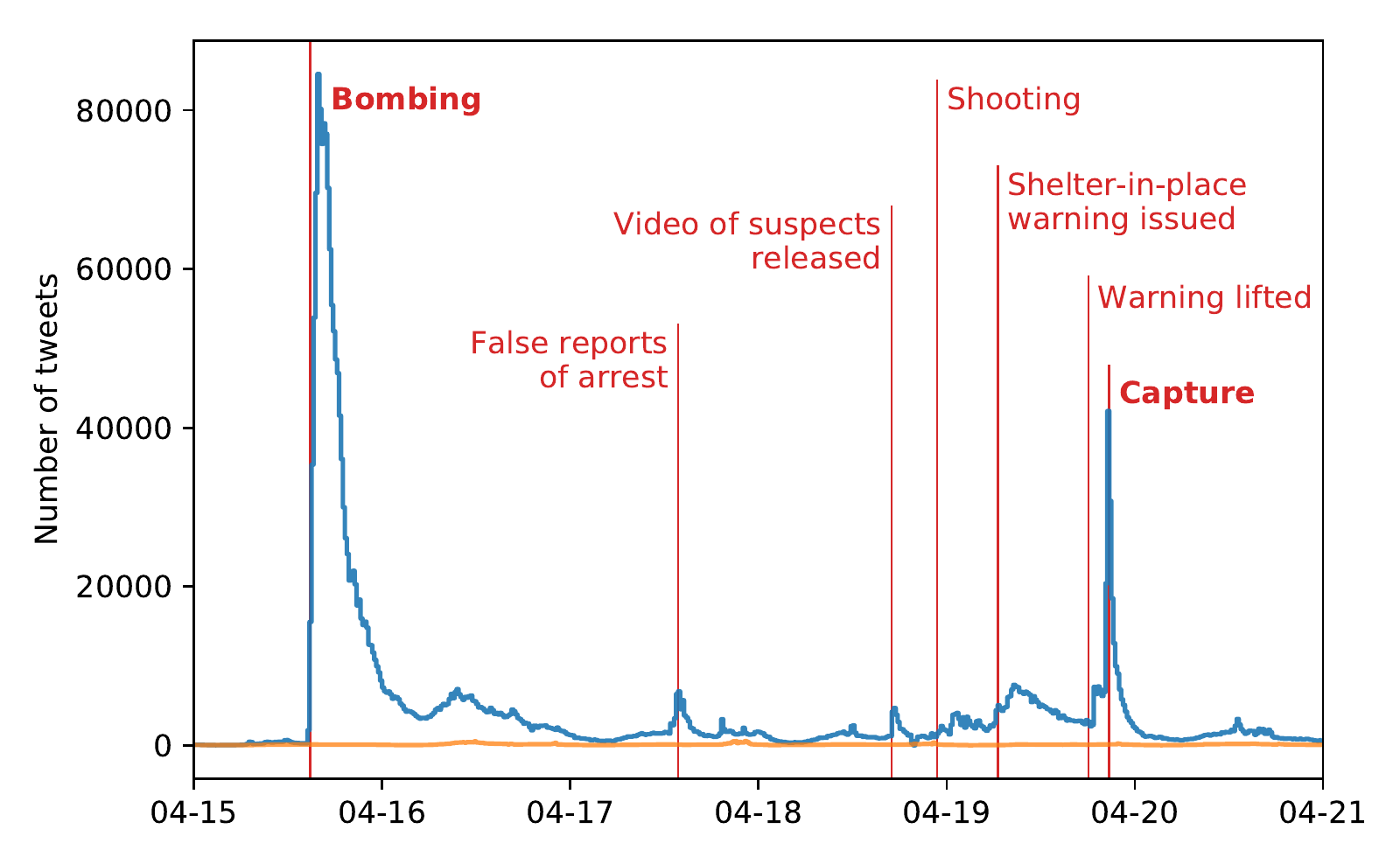}
\caption{Closeup on the Boston Marathon Bombing time series shown in Fig.~\ref{fig:numBostonTweets}.
The exogenous ``spiking'' pattern of both major events is clear, as are multiple other events occurring in the interim period between the bombing itself and the capture of the Dzhokhar Tsarnaev.
}
\label{fig:numBostonTweets_tight}
\end{figure}

The events and rumors surrounding the Boston Marathon Bombing and how they unfolded online provide a useful case study for information and rumor spreading. 
Here we studied Twitter activity during and after the Boston Marathon Bombing using data captured from the ``Gardenhose'' feed, which captures a random 10\% of all public Twitter activities.
Figure~\ref{fig:numBostonTweets} shows the volume of tweets (and retweets) over time containing ``boston'' (case-insensitive). 
Two strong spikes are present, coinciding with the bombing itself and Dzhokhar Tsarnaev's capture. For comparison, we also determined the number of tweets containing ``boston'' one year prior, and superimposed the two time series. Before the bombing these time series line up very well, demonstrating much year-over-year regularity. When the bombing occurs, the volume of related tweets increases by a factor of approximately 800.
The increased volume persisted for the rest of April.
A closeup of the event period itself (Fig.~\ref{fig:numBostonTweets_tight}) showed how well events surrounding the bombing are mirrored in the Twitter discussion.

\begin{figure}[t]
\sidecaption[t]
{\includegraphics[width=0.63\textwidth,trim=10 60 10 60,clip=true]{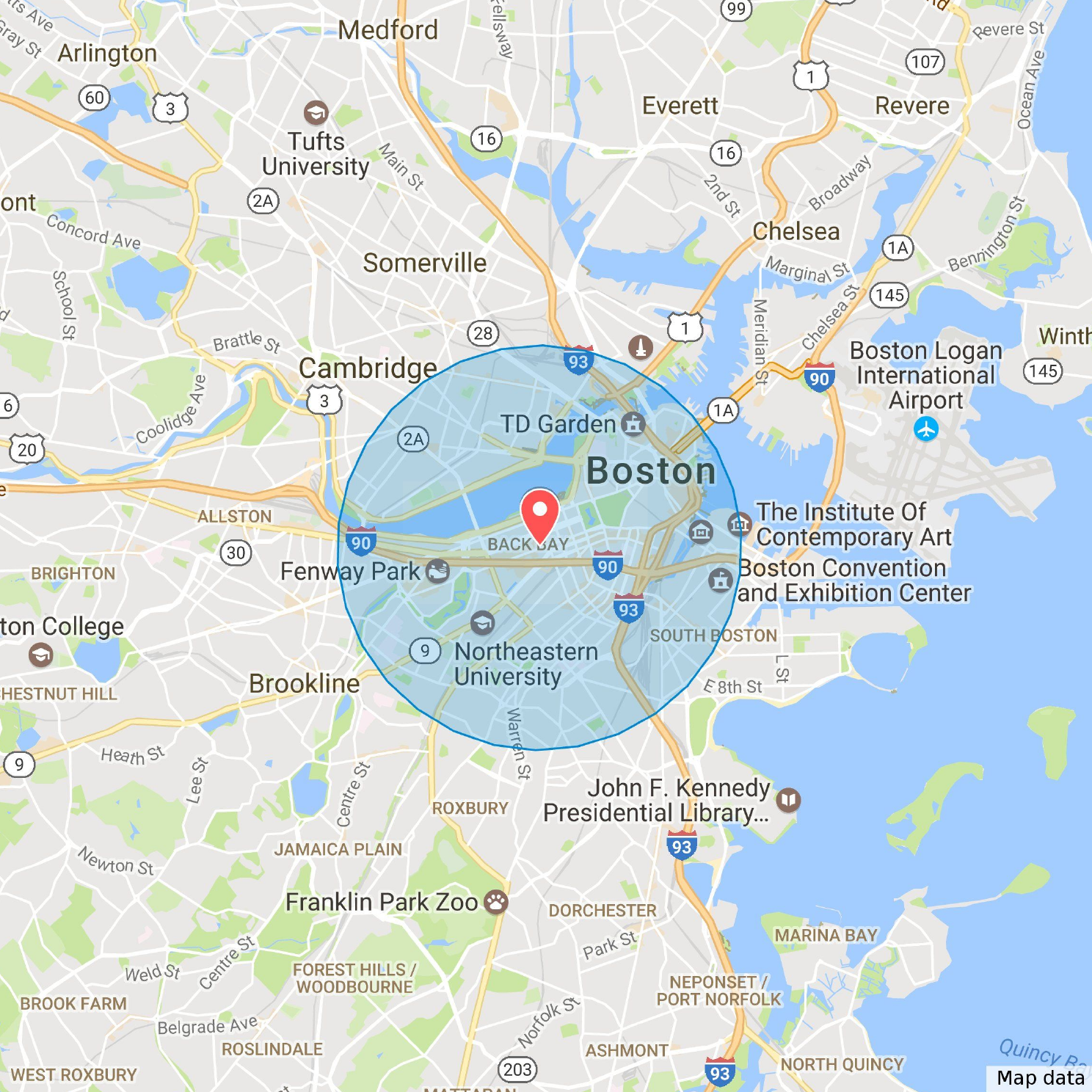}}
\caption{The blast site (marker) and selection area (circle) for the Boston Marathon Bombing. 
The circle forming the event region has radius 3 km. 
Users who post tweets from within the event region during the 5-hour period following the blast comprise the $G_0$ population.}
\label{fig:bostonBombingMap}
\end{figure}

\begin{figure}
\includegraphics[width=0.85\textwidth]{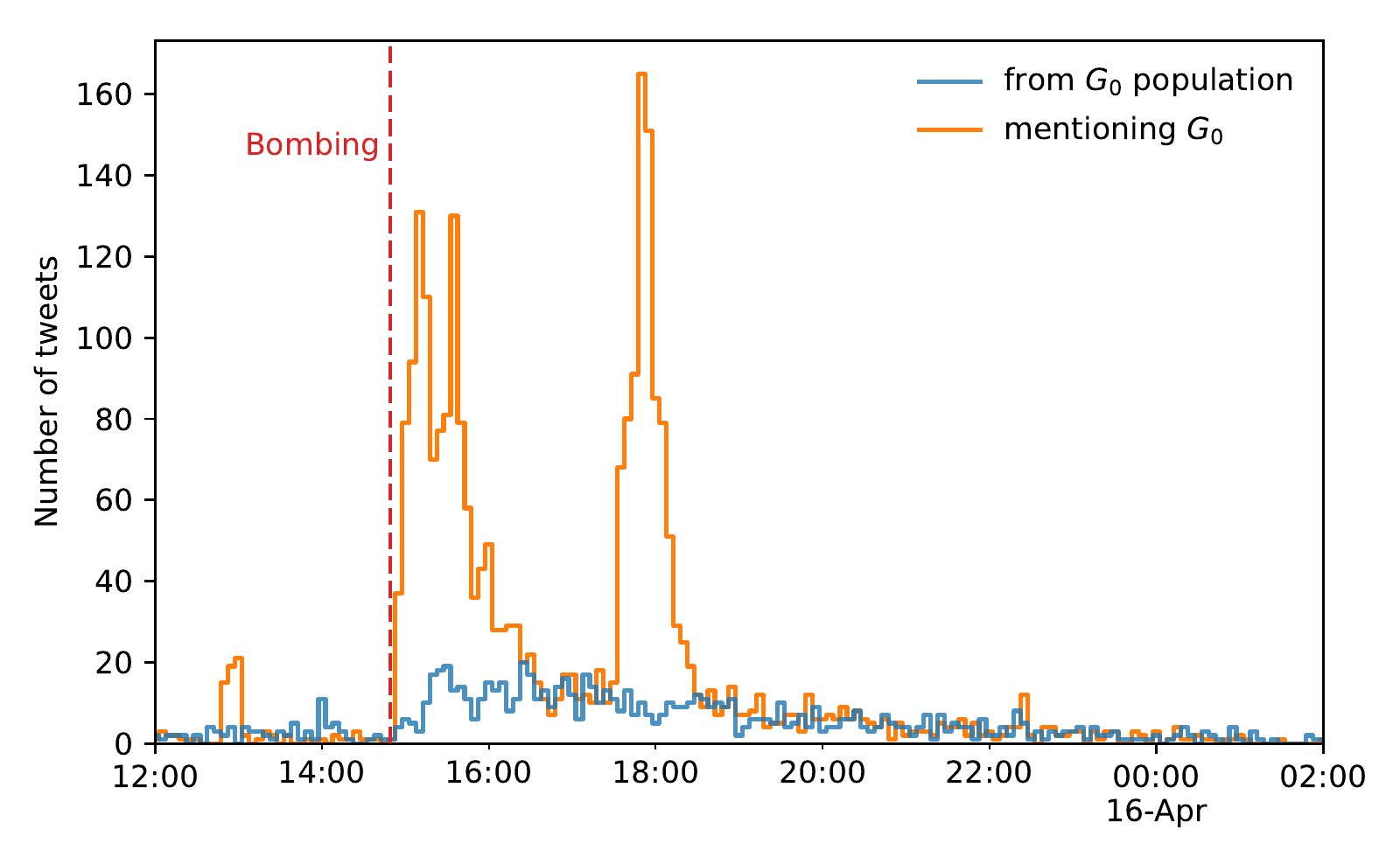}
\caption{Volume of tweets posted by members of the Boston Marathon Bombing $G_0$ population (Fig.~\ref{fig:bostonBombingMap}) and tweets at-mentioning members of the $G_0$ population.
The second spike beginning at approximately 17:30 is primarily due to a highly retweeted tweet reported by a member of the $G_0$ population about a possible suspect in custody. This event was later determined to be unrelated to the attacks.
}
\label{fig:g0_g1_counts}
\end{figure}

We next considered all geotagged tweets occurring within 3 km of the Boston Marathon Bombing blast site (Fig.~\ref{fig:bostonBombingMap}) during the 5-hour period immediately after the bombing occurred. The authors of these tweets form a population called $G_0$, those active ``tweeters'' in the vicinity of the event.
We then re-scanned the Gardenhose feed, capturing all the tweets posted by individuals within $G_0$ and all tweets which \textbf{\textit{mention}} individuals within $G_0$. 
Mentions (or ``at-mentions'') are a Twitter-specific term for posted tweets which contain the usernames of other Twitter users and are used to focus discussions and alert participants to online conversations; we used the mentioned usernames which Twitter extracted and provided as part of the Gardenhose feed.
Time series of tweet activity for $G_0$ individuals and mentions of $G_0$ individuals are showed in Fig.~\ref{fig:g0_g1_counts}.

Both time series show elevated activity levels in the aftermath of the bombing. In fact, the selection criterion for the $G_0$ population forces the time series to display a higher activity level, as that time series is now conditioned on the fact that tweets were posted after the bombing~\cite{bagrowDisaster2011pone}. 
Beyond this, we make two observations:
\begin{enumerate}
\item
The spike in mentions of $G_0$ individuals occurs more quickly than the spike in direct $G_0$ activity. This implies that Twitter is not being used to get information out of the event area as much as it is being used in parallel with other media such as news reports. Perhaps these tweets are people trying to reach social ties within the event region, although the nature of public Twitter activity makes it more likely that these are news reports and other media and government organizations. 

\item A very strong second spike in mentions is apparent several hours after the bombing. Inspecting tweets posted at this time showed that this is due to one highly viral (heavily ``retweeted'') tweet reporting the arrest of an individual as witnessed by a member of $G_0$. This second spike also peaked at a higher volume than the original spike, although it died out more quickly. This implies that the Twitter audience was primed to forward information during the immediate aftermath of the bombing, and the virality of any related content was much stronger.
An emergency event primes the audience of social media for rumoring and other information propagation.

\end{enumerate}

Taken together, the tragic Boston Marathon Bombing provides an exemplar case study for analyzing the interplay between human dynamics, information and misinformation spread, and communication media and social media.

\section{Mobile phone activity during emergencies}
\label{sec:mobilephonesemergencies}

Mobile phone datasets complement social media data for studying emergencies and disasters in many ways.
Mobile phones are generally more established in various regions of the world, having a longer history of use and higher levels of adoption, and providing years worth of extra historical records and large population samples. Mobile phone activity, especially voice calling, also lacks the broadcast nature of social media, acting instead as a direct communication channel. 
This direct communication means phone activity captures something very different than social media activity.

\begin{figure}[t]
{\includegraphics[trim=0 118 345 0,clip=true]{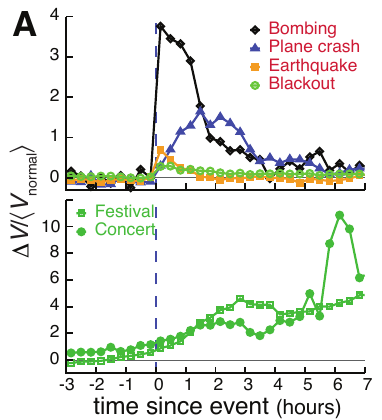}}
{\includegraphics[trim=0 0 345 122,clip=true]{figures/plosone_summaryFig}}
\caption{{Temporal and spatial response during emergencies, as measured from the mobile phone records
of a large service provider in a western European country.
    \lett{A} %
    The time dependence of call volume $V(t)$ (voice and text) after four
    emergencies and two non-emergencies. We plot the relative change in call
    volume $\Delta V/ \normal{V}$, where $\Delta V = \event{V}-\normal{V}$,
    $\event{V}$ is the call volume on the day of the event and $\normal{V}$ is
    the average call volume during the same period of the week.
    \lett{B} %
    The total change in call volume between two,  two-hour periods before and after a
    plane crash, as a function of distance $r$ from the epicenter of the crash.
    Following the event, we see an approximately exponential decay $\Delta V
    \sim \exp(-r / \rc)$ characterized by decay rate $\rc$.
    (Figure adapted from Bagrow \emph{et al.}~\cite{bagrowDisaster2011pone}.)
}}
\label{fig:plosOneTemporalSpatial}
\end{figure}

\begin{figure}
{\includegraphics[width=\textwidth,trim=123 0 0 0,clip=true]{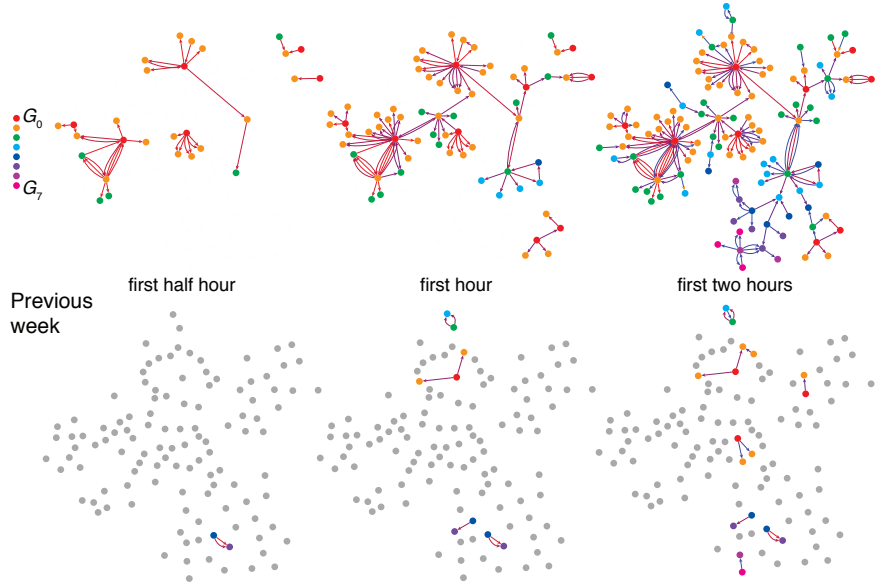}}
\caption{Part of the contact network formed between mobile phone users in the wake of the European bombing. 
Nodes are colored by group, with $G_0$ representing phone users calling from the event region, $G_1$ the recipients of those calls, etc. 
As time goes by more users are contacted as information propagates. 
Those same users make little contact during a corresponding time period the week before.
These snapshots show the social spreading one can observe from mobile phone data.
(Figure adapted from Bagrow \emph{et al.}~\cite{bagrowDisaster2011pone}.)
}
\label{fig:plosOneBombingContactGraph}
\end{figure}

\begin{figure}
\sidecaption[t]
{\includegraphics[width=0.645\textwidth,trim=6 5 4 6,clip=true]{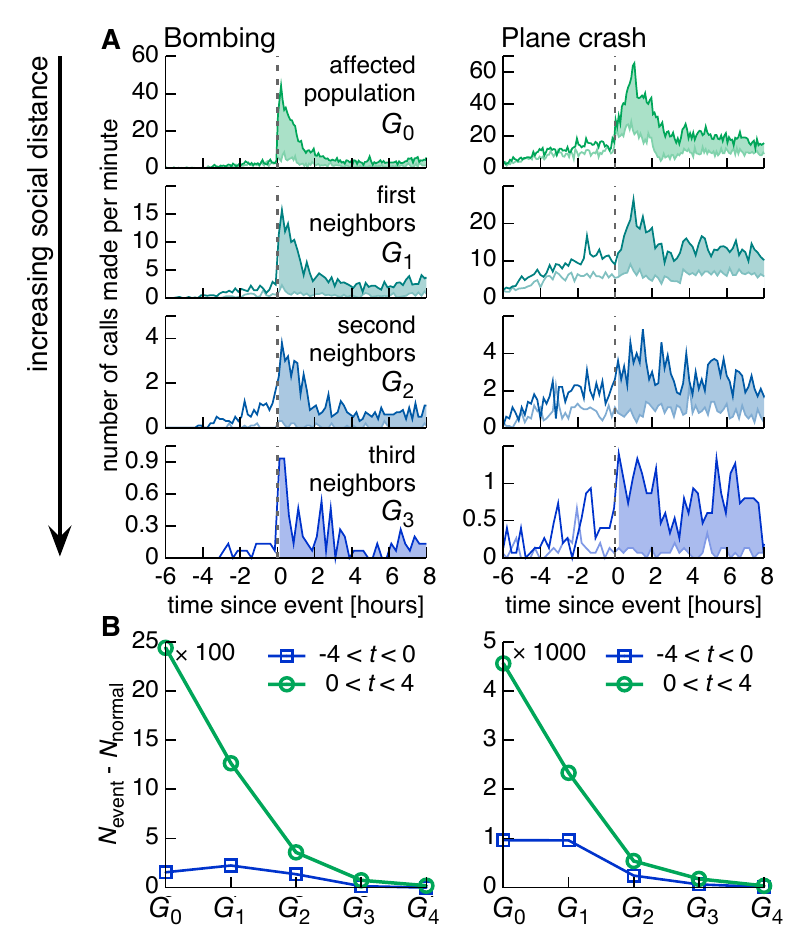}}
\caption{%
Social spread of activity following emergencies, as measured from mobile phone records. 
The most serious events show strong propagation across the contact network. 
\lett{A} %
Time series of call volume before and after the event, for each population $G_0, \ldots, G_3$. The shaded regions denote the extra or anomalous call volume from that population compared with their activity the week prior.
\lett{B} %
The total difference in call volume compared to the prior week during time periods before and after the event, for each $G_i$.
The two events and their $G_i$ populations are those studied from Bagrow, Wang, and Barab\'asi~\cite{bagrowDisaster2011pone}.
}
\label{fig:plosOneSocialSpread}
\end{figure}

\begin{figure}
\sidecaption[t]
{\includegraphics[width=0.645\textwidth]{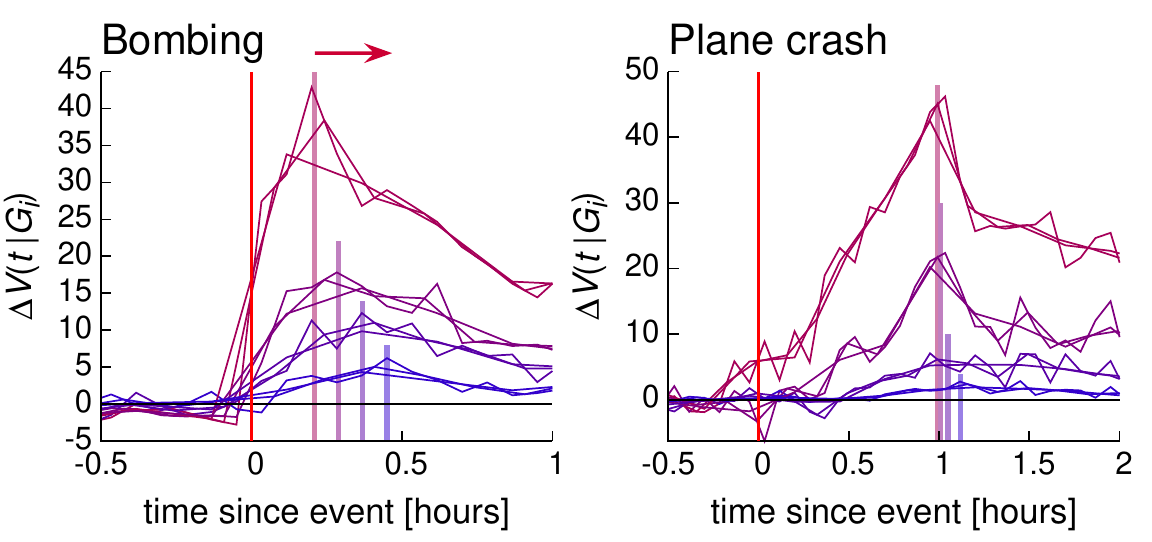}}
\caption{Outward social information spread is most evident for the bombing.
Different time series curves for each of $G_0, \ldots G_3$ correspond to 5-, 10-, and 15-minute time bins, intended to smooth the curves. Vertical marks denote the approximate peaks of each time series.
}
\label{fig:social_new_redux}
\end{figure}

In an earlier work, we studied activity levels in the wake of multiple emergency events using mobile phone records from a phone provider in a western European country~\cite{bagrowDisaster2011pone,gao2014quantifying}. 
These events included a bombing, a plane crash, and more. 
We found that the rapid spike in calls immediately following the emergency (Fig.~\ref{fig:plosOneTemporalSpatial}A) was spatially localized (Fig.~\ref{fig:plosOneTemporalSpatial}B), but rapidly propagated socially for the most serious events (Figs.~\ref{fig:plosOneBombingContactGraph}, \ref{fig:plosOneSocialSpread}, and \ref{fig:social_new_redux}).
This social propagation was measured from the time series of call activity for different populations of mobile phone users: 
$G_0$, the eyewitness group, calling from the direct vicinity of the event; $G_1$, those individuals who receive calls from members of the $G_0$ group during the time period of the event; $G_2$, etc. As $i$ increases, the group $G_i$ becomes more social distant from the event itself.

The bombing in western Europe provides the most clear evidence for social information spreading based on the time series of call activity (Fig.~\ref{fig:social_new_redux}A). Here we denote on the figure the times of the peaks of call volume for each group $G_i$ using vertical bars. A temporal ordering is clearly evident for the bombing with the peak cascading through the populations over an approximately 20-minute period (denoted by the horizontal arrow). The plane crash (Fig.~\ref{fig:social_new_redux}B) does not show such clear temporal ordering of the peak. This may be due to the fact that news media were covering the crash and that social ties, particularly members of $G_1$,  were likely to already be aware that their contacts were traveling that day. This underscores the different natures of equally unexpected emergency events.

This social spread of information outward from the wake of an emergency is intuitive, but we did not observe it in the Twitter data following the Boston Marathon Bombing. Indeed, in that case, unaffected individuals mentioning affected individuals spiked in activity before those who tweeted within the Boston Marathon Bombing event region (Fig.~\ref{fig:g0_g1_counts}). This underscores the strong influence the communication channel has: for mobile phones, it is a direct communication channel and that limited scope requires a $G_0$ individual to carefully choose who to contact, but for Twitter it is a secondary broadcast meant to update many followers. Those followers are likely to be less socially close than contacts reached by mobile phone call, and it is probable (though definitely not certain) that a $G_0$ individual will turn to phone calls first in the wake of an event, and then only later begin to use social media. And of course, mobile phones are not confounded by news organizations, government entities, and journalists the way social media are.

\section{Detecting anomalous events}
\label{sec:miningemergencies}

Given that emergencies and disaster events are useful for understanding and observing how information spreads in context, it is also worth understanding how rare these events are.
To estimate the rate of emergencies and non-emergency events (collectively called \emph{anomalies}) in modern datasets, as well as provide an example of the types of analysis now possible, here we introduce and apply an anomaly detection method to a country-wide mobile phone dataset, and use several basic descriptive measures on the identified anomalies to characterize their features.
Such algorithms can in principle be used to detect the onset of an emergency event in real-time. This is a crucial application for first responders. However, here our focus is only on discovering anomalous events after they occur, so that they may be retroactively studied.

\subsection{Detecting Anomalies}

We implement a basic event detection algorithm and apply it to the six-month time series' of call volume taken from the mobile phone call detail records.
This algorithm exploits the periodicity and recurrent nature of mobile phone activity patterns and performs well with noisy data.  (A more advanced method, the Markov-modulated Poisson process~\cite{ihler2007learning}, proved inadequate for this dataset.)

We first pre-processed the data.  
To help with heterogeneous tower densities, we began by dividing the country into equally-spaced squares of size $1\times1$ km (one can also use $10 \times 10$ km grids).  
All cell towers sharing a grid space were merged so that the total volume $V_\mathbf{x}(t)$ of phone calls at grid space $\mathbf{x}$ is the sum of the call volumes of all towers within $\mathbf{x}$.
Grid spaces that do not contain cell towers were neglected. 
We now refer to each square grid space as a location. 
Since our goal is to find events that can yield good statistics, we ignored locations that are mostly unoccupied by only considering locations that average at least 1 phone call per minute over the entire six-month period.
These time series are then binned into ten-minute intervals so that their total length is $6\times24\times7\times W$ (covering $W$ weeks).

The algorithm uses two calculations to flag \emph{runs} of suspiciously high call volume for each time series, where a run is a time period denoted by a start time and a stop time).
Runs that overlap in time (or nearly overlap) are merged\footnote{Specifically, two adjacent runs of suspicious time periods are merged if they overlap in time or they are separated by less than four time bins and at least one of the two runs is longer than four time bins}.   
After mergers,
a run must have at least one time bin flagged as suspicious by both calculations and have a duration of at least 5 time bins to be considered an anomaly.

The two calculations to flag runs of suspiciously high activity use the \emph{variance} (Sec.~\ref{subsubsec:variancecalcdetectinganomalies}) and
the \emph{recurrence} (Sec.~\ref{subsubsec:recurrencecalcdetectinganomalies}) of the $V_\mathbf{x}(t)$.

\subsubsection{Variance calculation}
\label{subsubsec:variancecalcdetectinganomalies}

Each location's time series $V_\mathbf{x}(t)$ is copied $W$ times, with each
copy circularly rotated by one week from the previous copy. 
Now each 10-minute bin $t$ can be compared to all the other bins that occur at that same time of the week.  
Dropping the location index, let us denote $V(t)$ as the original time series, $\left<V_\mathrm{shifted}(t)\right>$ as the average of element $t$ over the $W$ rotated copies, and $\sigma\left(V_\mathrm{shifted}(t)\right)$ as the standard deviation of the $W$ rotated copies.  
Now we construct a new vector $Z(t)$,
\begin{equation}
    Z(t) = \frac{V(t) - \left<V_\mathrm{shifted}(t)\right>}{\sigma\left(V_\mathrm{shifted}(t)\right)}.
\end{equation}
Finally, we flag as suspicious those contiguous times $t_s \in [t_\mathrm{start},t_\mathrm{stop}]$ where $Z(t_s) > Z_\mathrm{thr}$ for all $t_s$.
In other words, a suspicious event's $t_\mathrm{start}$ and $t_\mathrm{stop} > t_\mathrm{start}$ are determined by those times $t$ where $Z(t)$ crosses and remains above $Z_\mathrm{thr}$.
Events where only a single time bin was flagged ($t_\mathrm{start} = t_\mathrm{stop}$) are ignored.
For this work we use $Z_\mathrm{thr} = 2.5$.

\subsubsection{Recurrence calculation}
\label{subsubsec:recurrencecalcdetectinganomalies}

Take the original time series $V(t)$ (suppressing location index) and rotate it by $10\tau$ minutes ($\tau$ elements). 
One can construct a recurrence or Poincar\'e plot by plotting the original time series $V(t)$ against the rotated series $V(t+\tau)$.

\begin{figure}[!t]
\centering
\includegraphics[width=0.8\textwidth]{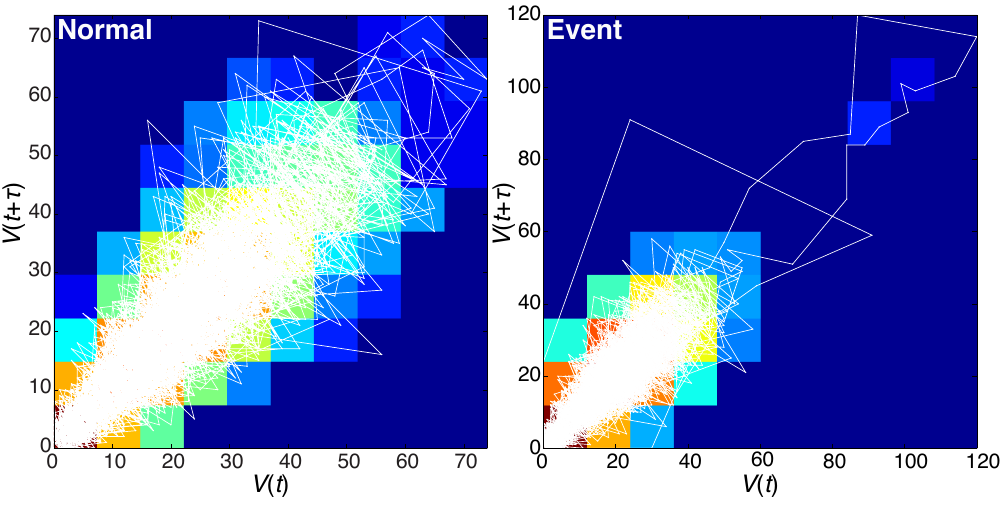}
\caption{Recurrence plots of $V(t)$ for $\tau=10$ minutes.  
Colored squares indicate the log of the probability for a randomly chosen point to fall in that bin.
The run of points in an otherwise unoccupied region in the upper-right corner of the right plot indicates a persistent deviation from the expected recurrence and is flagged as suspicious.
\label{fig:recurrenceplot}}
\end{figure}

If the time series is periodic, the plot will trace out a
circular trajectory (Fig.~\ref{fig:recurrenceplot}).  
Deviations away from the normal pattern will appear as regions of the phase space with relatively few points.  
To detect these regions we bin the phase space into squares of size $20
\times 20$ minutes.
The probability for a bin to contain a randomly chosen point is estimated as the fraction of points that fall within that bin.  
We flag points as suspicious if the probability to be in that bin is less than $1/(24\times7\times W)$. 
For this work we use two rotations, one being 10 minutes ($\tau=1$), which primarily looks for sudden changes in activity, and the other being 1 week ($\tau=6\times24\times7$), which focuses on changes from
the weekly periodicity.  
A point in time is suspicious if it is flagged in either recurrence plot.

\subsubsection{Results}

We applied the algorithm defined above to six months of mobile phone data records, and detected a total of 340 call anomalies.
This corresponds to an average of 1.8 anomalies per day.
Therefore, we conclude that researchers with access to years or decades of activity data may have records of hundreds or even thousands of small-, medium-, and large-scale anomalies to study.
While many of the 340 events detected are not emergencies\footnote{We inspected the anomalies manually and determined the origins of many of the events using Google News, but cannot share this information as it will reveal the country of origin of the data, breaking our non-disclosure agreement with the mobile phone provider.}, even non-emergency events provide a view into social activity and information spreading that is not available when one is limited to studying normal periods of activity.

\subsection{Characterizing detected events}

\begin{figure}[!t]
    \includegraphics[width=\textwidth]{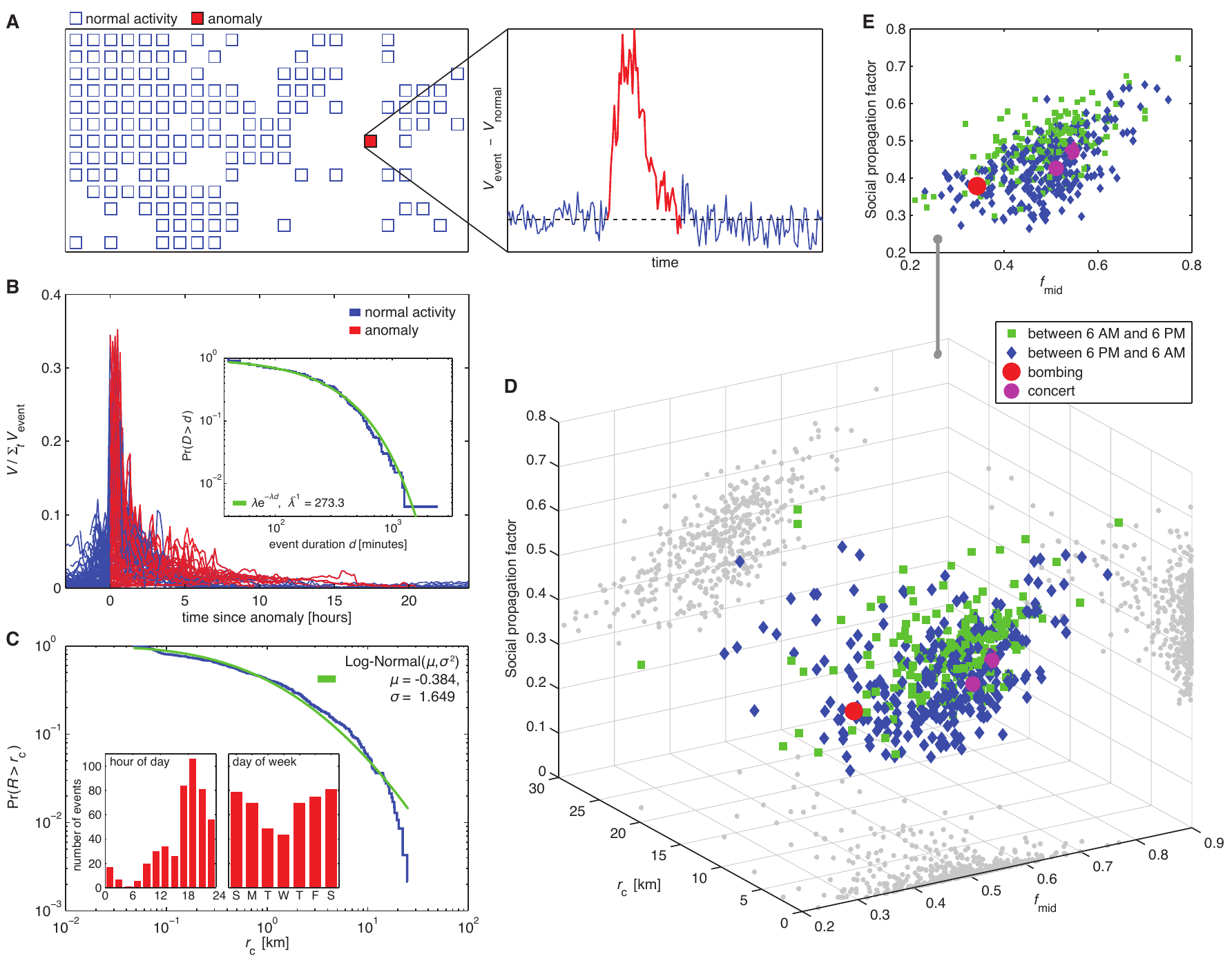}
    \caption{
Systematic anomaly detection to estimate the rate of anomalies captured by mobile phone data records. 
\lett{A} The full country is divided into 1 km $\times$ 1 km grids. 
Assigned to each grid space is a six-month time series corresponding to activity from mobile phone towers within that space.
A composite detection algorithm, exploiting daily, weekly, and seasonal periodicities in call activity, is then used to flag anomalous call periods (highlighted). 
The final result is a corpus of 340 anomalies. 
The bombing and several known concerts occurred during this six-month period, and were successfully identified.
\lett{B} 
Time series for some anomalies, scaled so that the total activity during the anomaly is unity. 
\lett{inset}
The distribution of anomaly durations is approximately exponential, with an average duration of 273 minutes. 
\lett{C} 
The distribution of characteristic spatial distances $\rc$. 
The average $\avg{\rc} = 2.33$ km corresponds well to the events studied by Bagrow \emph{et al.}~\cite{bagrowDisaster2011pone}. 
A log-normal distribution is shown for comparison. 
\lett{inset} 
The distribution of anomaly start times, as a function of time of day and day of week. 
Fifty one percent of anomalies occur between 6PM and midnight. 
\lett{D-E}
For each anomaly, we plot: the midpoint fraction $\fmid$, the time it takes for half the anomalous call activity to occur; $\rc$; and the social propagation factor, measuring how rapidly the anomaly propagates through the social network. 
We see that propagation rates are independent of $\rc$ and that the bombing shows faster propagation than the concerts. 
Interestingly, events that occur during the day tend to show slower social propagation than events that begin during nighttime hours.
    }
\label{fig:bin_grid_toon}
\end{figure}

After identifying a call anomaly using the above procedure (Fig.~\ref{fig:bin_grid_toon}A), we can characterize its temporal, spatial, and social properties:

\begin{description}

\item[\bf Temporal] The temporal nature of an event can be captured by how quickly it peaks. However, the time of the peak itself is often difficult to measure accurately from a noisy time series and may be influenced by any binning of the time series. 
Instead, we measure $\fmid$, the midpoint fraction, defined as the fraction of time it takes for half of the total anomalous call activity to occur. When there is a sharp spike in call volume, as shown in the red curve in Fig.~\ref{fig:bin_grid_toon}A, $\fmid$ will be low.
Specifically, $\fmid = (t_\mathrm{mid} - \tstart)/(\tstop - \tstart)$, where $t_\mathrm{mid}$ is defined such that
\begin{equation}
\int_{\tstart}^{t_\mathrm{mid}} \left(\event{V}(t) - \normal{V}(t)\right) dt = \frac{1}{2} \int_{\tstart}^{\tstop} \left(\event{V(t)} - \normal{V}(t)\right) dt.
\end{equation}

\item[\bf Spatial] How much an event's call anomaly is localized spatially around the detected epicenter can be captured by its characteristic spatial decay rate $\rc$.
We measure this by integrating the anomalous call activity in concentric rings of radius $r$ around the event epicenter, and fit an exponential function, i.e., $\Delta V(r) \sim \exp(-r / \rc)$. The spatial decay rate tells us whether the event is sharply peaked at a location (small $\rc$) or spreads broadly over space (large $\rc$).

\item[\bf Social] The social spread of a call anomaly can be measured by analyzing statistics of the time series of calls made by populations $G_0, G_1, \ldots$. These populations capture those directly affected by the event ($G_0$), those who receive calls from $G_0$ but are not themselves members of $G_0$ ($G_1$), the recipients of call from $G_1$ members not in $G_1$ or $G_0$ ($G_2$), etc. To capture how quickly the call anomaly spreads through these populations, we define the \textit{social propagation factor} as simply the midpoint fraction of the time series of anomalous call volume for each population $G_i$, averaged over $G_i$.

\end{description}

In Fig.~\ref{fig:bin_grid_toon}B we present re-scaled time series of the call activity during the detected anomalies, compared with activity under non-anomaly circumstances. Most anomalies are short in duration, lasting under 2 hours, although a few were detected lasting over 20 hours. Likewise, most anomalies were spatially localized (Fig.~\ref{fig:bin_grid_toon}C). Most anomalies occurred after 18:00 local time, although most phone activity also occurred after 18:00 so this observation may be a simple confound.
Weekends were more likely to contain anomalies, with Wednesday being the weekday having the fewest detected anomalies. 

The speed of the event, measured by how quickly the localized call anomaly peaks, correlates well with the social propagation factor measuring how quickly the call anomaly peaks within the social populations $G_0, G_1, \ldots$ (Fig.~\ref{fig:bin_grid_toon}D,E).
This relationship is roughly independent of spatial localization as measured by $\rc$ (Fig.~\ref{fig:bin_grid_toon}D).
We also observe that social propagation is slower for daytime events (those occurring between 06:00 and 18:00 local time), regardless of $\fmid$ itself (Fig.~\ref{fig:bin_grid_toon}D,E). 
For an event to occur in the middle of the night and have a strong social propagation factor is good evidence that it is an emergency or disaster.

\subsubsection{Principal Component characterization}

\begin{figure}[t!]
    \centerline{\includegraphics[width=0.5\textwidth,trim=0 0 410 0, clip=true]{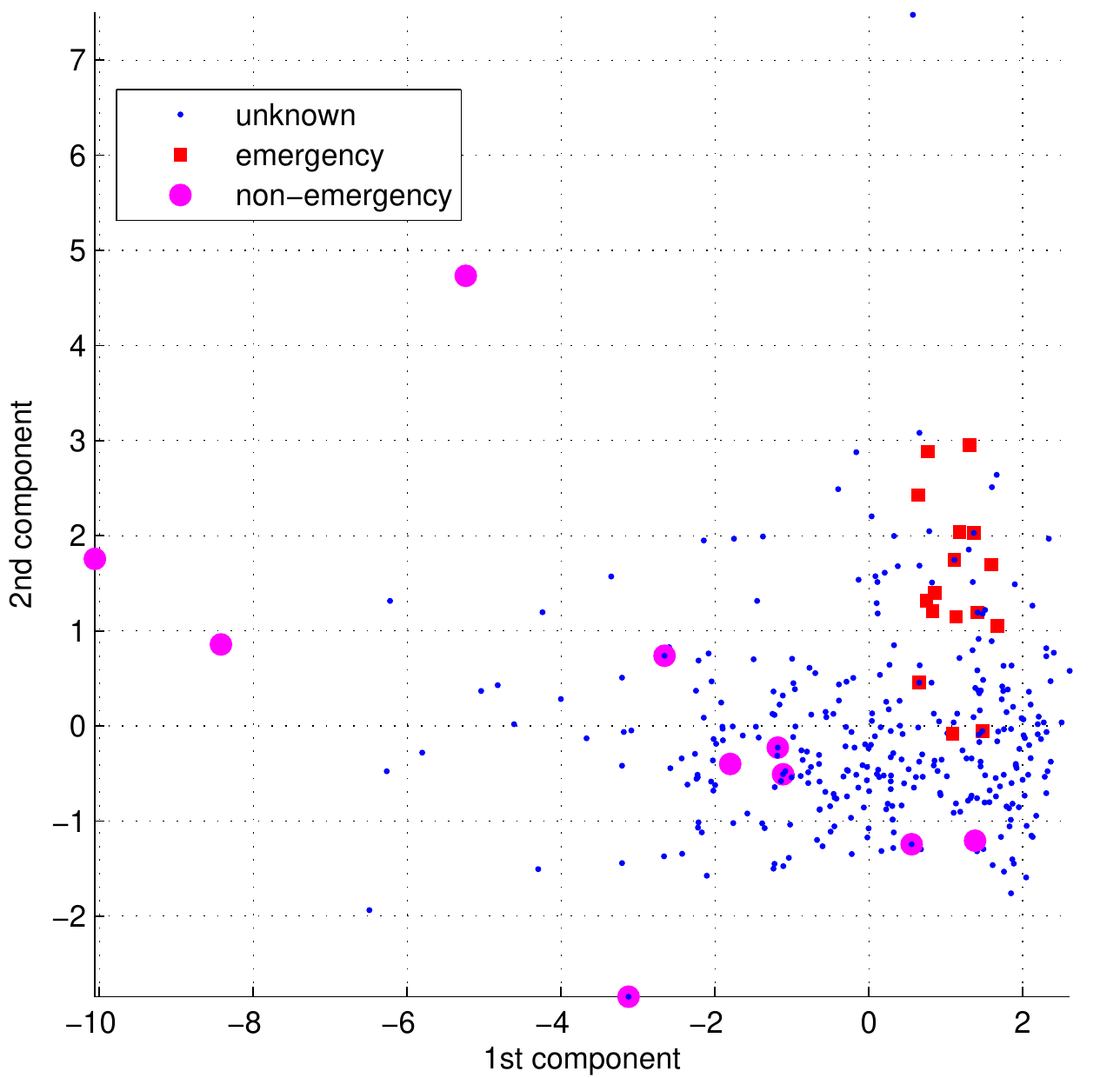}
                \includegraphics[width=0.5\textwidth,trim=0 0 410 0, clip=true]{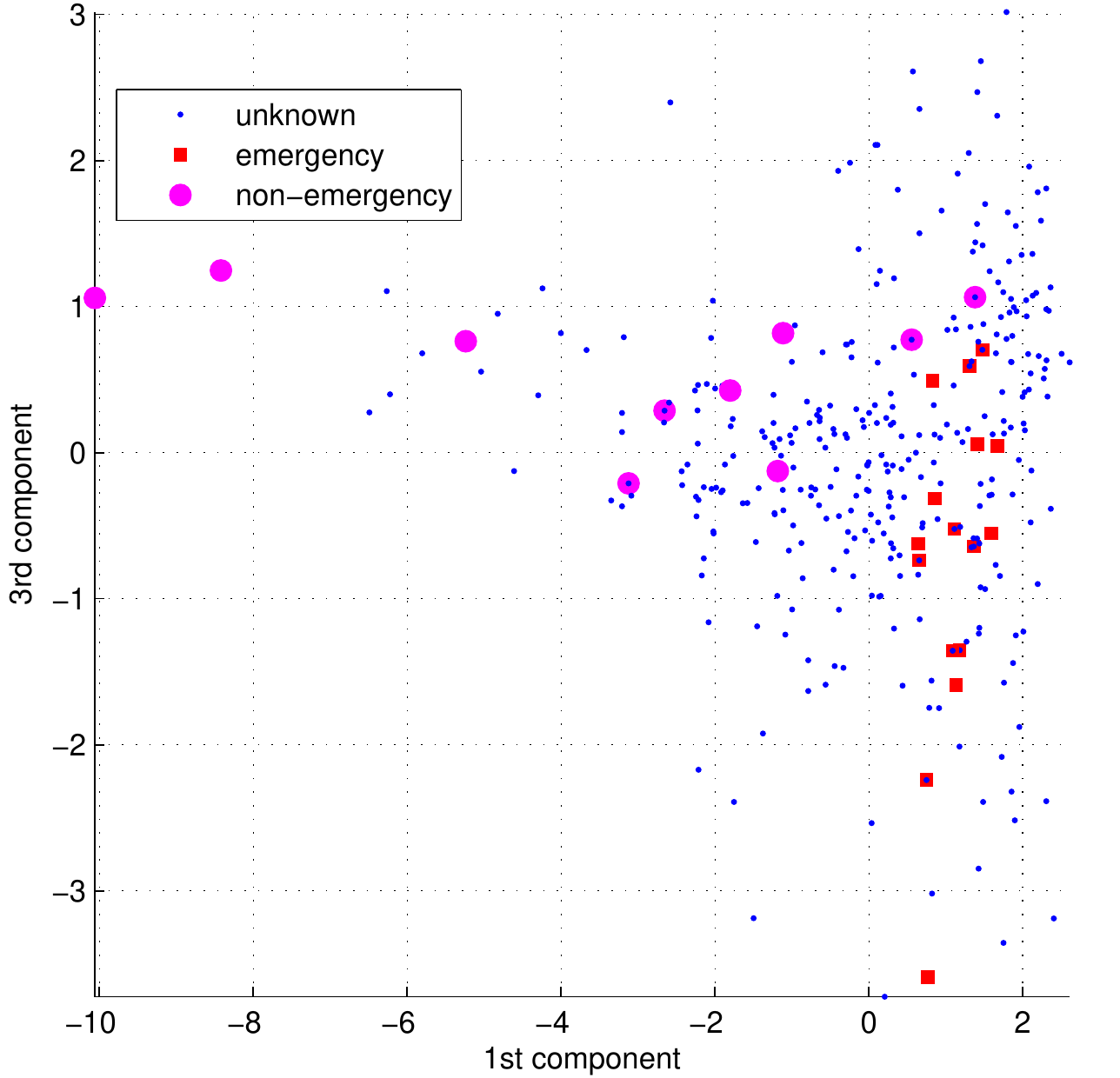}}
    \caption{Principal component analysis for the anomalies detected in the mobile phone records.
    Anomalies were manually inspected and many were found to correspond to known emergency and non-emergency events.
    Emergency events showed distinct clustering, particularly in the first two principal components.
    \label{fig:pca_ents}}
\end{figure}

Lastly, we performed a principal component analysis (PCA)~\cite{jolliffe2002principal} on the 340 detected events (including events that were found by manual inspection to correspond to those studied by Bagrow \emph{et al.}~\cite{bagrowDisaster2011pone}). 
Nine measurements (or features) were determined for each event:
the spatial size of the event $\rc$;
the speed at which the event occurs $\fmid$;
the time of day; 
event duration; 
total number of calls;
affected population size $\left|G_0\right|$;
the ``social decay rate,'' the ratio of the total number of calls made by population $G_i$ vs.\ $G_{i-1}$ averaged over i;
the z-score for the total number of anomalous calls placed by population $G_i$, averaged over $i$;
and weighted social distance $\sum_i i \times V_\mathrm{total}(G_i) / \sum_i V_\mathrm{total}(G_i)$.
These measures are intended to capture many different aspects of the call anomalies, and more can in principle by used, under the assumption that PCA will ``net out'' the most relevant linear combinations of these features.
A $340 \times 9$ data matrix is then constructed.
The first three principal components are shown here (Fig.~\ref{fig:pca_ents}).
We found a clustering of known emergencies, with known non-threatening events appearing mostly as outliers.
The clustering of emergencies is evidence that the measures introduced here can be used to categorize events 
without additional information.

\section{Discussion}
\label{sec:discussion}

In this chapter, we discussed how to measure information or activity spreading through a social system in the wake of an emergency, disaster or other anomalous event. 
Such emergency events act as ``found experiments,'' providing researchers with new contexts and windows on the underlying social system.
We presented a case study of information spread on Twitter following the Boston Marathon Bombing, and described measures of social spreading within a western European country captured from mobile phone records. 
Mobile phone data are limited in scope---lacking, for example, contextual details such as the text information available in social media---and are generally less freely available to researchers.
But mobile phone datasets strongly complement other data such as those taken from Twitter because phone calls and text messages represent one-on-one, direct communication and are not confounded by broadcast effects and news media the way Twitter is.

Comparing the spreading dynamics on Twitter surrounding the Boston Marathon Bombing with the western European Bombing captured from mobile phone records underscores how different these communication media are, both in who uses these media and what is expected from these media. 
Researchers must account for these differences when studying and comparing across media. 
Even within a single type of media there may be great differences: a photo-oriented platform like SnapChat may present vastly different dynamics than a microblogging platform like Twitter or a chat platform like WhatsApp or Facebook Messenger.
Further, spreading in a single platform does not take place in isolation: the dynamics of Twitter users following the Boston Marathon Bombing are strongly influenced by information they (or their social ties) receive from traditional, broadcast media.

As communication services continue to evolve, and online activity continues to adapt to new services, researchers will be confronted with both technical challenges to overcome but also a wealth of new opportunities brought about by new data.
Recent advances in machine learning and artificial intelligence may prove fruitful here, for example. Deep learning for computer vision may soon allow researchers to better understand and analyze video feeds and imagery created by eyewitnesses of emergency events, particularly when those feeds are generated in large volumes, keeping pace with new smartphone video streaming services such as Facebook Live and Periscope.

\begin{acknowledgement}
We thank S. Lehmann and Y.-Y. Ahn for organizing this book and inviting us to contribute, C.~M.~Danforth for useful comments on the Twitter data,
and we
gratefully acknowledge the resources provided by the Vermont Advanced Computing Core. This material is based upon work supported by the National Science Foundation under Grant No. IIS-1447634.
\end{acknowledgement}


\end{document}